\documentclass[12pt]{article}
\usepackage{graphicx}

\newcommand\pubdate{\today}

\def\Title#1{\begin{center} {\Large #1 } \end{center}}
\def\Author#1{\begin{center}{ \sc #1} \end{center}}
\def\Address#1{\begin{center}{ \it #1} \end{center}}

\newcommand\pubblock{\rightline{\begin{tabular}{l}  \\ 
         \pubdate  \end{tabular}}}
\newenvironment{Abstract}{\begin{quotation}  }{\end{quotation}}
\newenvironment{Presented}{\begin{quotation} \begin{center} 
             PRESENTED AT\end{center}\bigskip 
      \begin{center}\begin{large}}{\end{large}\end{center} \end{quotation}}

\begin{document}
\begin{titlepage}
 \pubblock
\vfill
\Title{Transverse single spin asymmetry for very forward
neutron production in polarized $p+p$ collisions at $\sqrt{s} = 510$ GeV}
\vfill
\Author{Minho Kim}
\Address{RIKEN}
\vfill
\begin{Abstract}
In the high-energy $p+p$ collisions, the transverse single spin asymmetry for very
forward neutron production has been interpreted by an interference between 
$\pi$ (spin flip) and $a_1$ (spin non-flip) exchange with a non-zero phase shift.
The $\pi$ and $a_1$ exchange model
predicted the neutron asymmetry would increase in magnitude with
transverse momentum ($p_{\scriptsize{\textrm{T}}}$) 
in $p_{\scriptsize{\textrm{T}}} < 0.4$ GeV/$c$.
In June 2017, the RHICf experiment installed an electromagnetic calorimeter at the
zero-degree area of the STAR experiment at the Relativistic Heavy Ion Collider and
measured the neutron asymmetry in a wide
$p_{\scriptsize{\textrm{T}}}$ 
range of $0 < p_{\scriptsize{\textrm{T}}} < 1$ GeV/$c$ 
from polarized $p+p$ collisions at $\sqrt{s} = 510$ GeV.
The RHICf data
allows us to investigate the kinematic dependence of the neutron asymmetry
in detail, which not only
can test the $\pi$ and $a_1$ exchange model
in the higher $p_{\scriptsize{\textrm{T}}}$ range but also can study the $\sqrt{s}$
dependence by comparing with the previous measurements.
We present the preliminary result and analysis status of the
neutron asymmetry measured by the 
RHICf experiment.
In order to understand the RHICf result,
a theoretical calculation other than Reggeon exchange will also be discussed. 
\end{Abstract}
\vfill
\begin{Presented}
DIS2023: XXX International Workshop on Deep-Inelastic Scattering and
Related Subjects, \\
Michigan State University, USA, 27-31 March 2023 \\
     \includegraphics[width=9cm]{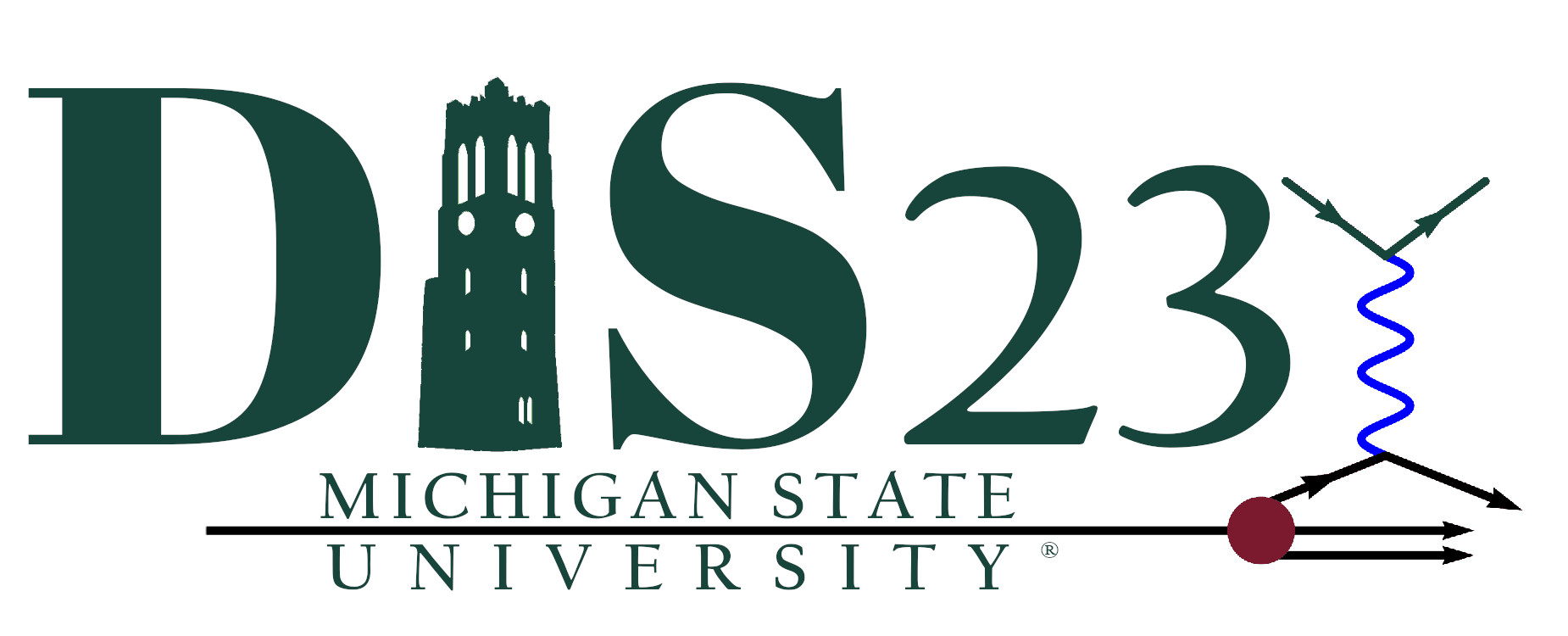}
\end{Presented}
\vfill
\end{titlepage}

\section{Introduction}
\quad \ \ In polarized $p+p$ collisions,
transverse single spin asymmetry ($A_{\scriptsize{\textrm{N}}}$) is defined by 
a left-right cross section asymmetry,
\begin{eqnarray*}
A_{\scriptsize{\textrm{N}}} = \frac{\sigma_{\scriptsize{\textrm{L}}}-
\sigma_{\scriptsize{\textrm{R}}}}{\sigma_{\scriptsize{\textrm{L}}}+
\sigma_{\scriptsize{\textrm{R}}}},
\label{eq:AN}
\end{eqnarray*}
where the $\sigma_{\scriptsize{\textrm{L}} (\scriptsize{\textrm{R}})}$ 
is the cross section of a specific particle which is produced
in the left (right) side with respect to the beam polarization.
Since a large $A_{\scriptsize{\textrm{N}}}$ 
for very forward ($\eta > 6$) neutron production
has been discovered by a polarimeter development experiment \cite{neuAN1},
it has been measured by the PHENIX experiment at three different collision 
energies, 62, 200, and 500 GeV \cite{neuAN2}.
One pion exchange (OPE) model
which had successfully described the 
cross section
of the neutron production introduced an interference between the spin flip
$\pi$ exchange and spin non-flip $a_1$ exchange with a relative phase shift
to reproduce the experimental data \cite{OPE5}. 
The $\pi$ and $a_1$ exchange model reproduced the data well predicting that
the $A_{\scriptsize{\textrm{N}}}$ increases in magnitude with 
transverse momentum ($p_{\scriptsize{\textrm{T}}}$) 
with little longitudinal momentum fraction ($x_{\scriptsize{\textrm{F}}}$) dependence. 
Recently, the neutron $A_{\scriptsize{\textrm{N}}}$ measured by the PHENIX experiment
at $\sqrt{s} = 200$ GeV
was unfolded to precisely study the kinematic dependence of the neutron 
$A_{\scriptsize{\textrm{N}}}$
and showed a consistency with the model prediction \cite{phenix}.
However, kinematic range of the experimental measurement and model calculation
was narrow compared to that of very forward neutron production.

The RHICf experiment \cite{rhicf} measured the $A_{\scriptsize{\textrm{N}}}$ for very forward
neutron production in a wide $p_{\scriptsize{\textrm{T}}}$ coverage of 
$0 .0 < p_{\scriptsize{\textrm{T}}} < 1.0$ GeV/$c$ from polarized $p+p$ collisions
at $\sqrt{s} = 510$ GeV.
The RHICf data allows us to test the validity of the $\pi$ and $a_1$ exchange model
in the higher $p_{\scriptsize{\textrm{T}}}$ region.
Since $\sqrt{s}$ of the RHICf experiment is different from those of previous 
measurements, one can also study the $\sqrt{s}$ dependence of the neutron
$A_{\scriptsize{\textrm{N}}}$.

\section{RHICf experiment}
\quad \ \ In June 2017,
The RHICf experiment measured the $A_{\scriptsize{\textrm{N}}}$ 
for very forward neutron production in
polarized $p+p$ collisions at $\sqrt{s} = 510$ GeV at the Relativistic Heavy Ion
Collider (RHIC).
An electromagnetic calorimeter (RHICf detector) \cite{rhicfdet}
was installed in front of the STAR
Zero-Degree Calorimeter (ZDC) \cite{zdc}, which was located 18 m away
from the beam interaction point.
We also installed a thin scintillator counter (FC) in front of the RHICf detector
to analyze the charged hadron background.
The RHICf detector consists of small and 
large sampling towers
with 20 and 40 mm dimensions, respectively.
Both towers are composed of 17 tungsten absorbers with 1.6 interaction length in total,
16 GSO plates for energy measurement,
and 4 layers of GSO bar hodoscope for position measurement.
Position and energy resolutions for 250 GeV neutron are about
1 mm and 39$\%$ when the incident position of the neutron is the center
of a tower.
A shower trigger which was operated when the energy deposits of three successive GSO
plate layers are larger than 45 MeV was used for the neutron measurement.
For more detailed detector configuration and performance, see Ref.~\cite{more1}, 
\cite{more2} and \cite{more3}.

We requested large $\beta^{*}$ value of 8 m at operation to make the
angular beam divergence small.
Corresponding luminosity at $\sqrt{s} = 510$ GeV
was $\sim$$10^{31}$ cm$^{-2}$s$^{-1}$ which was smaller
than that in usual RHIC operation.
We also requested a radial polarization which was 90$^{\circ}$-rotated
from that of usual RHIC operation to reach the maximal $p_{\scriptsize{\textrm{T}}}$
range by moving
the detector vertically.
We took the data during about 28 hours with three detector positions where the beam
headed the center of large tower, center of small tower, and 24 mm below the center
of small tower.

\section{Data analysis}
\quad \ \ Basically, only the shower triggered events were analyzed to avoid any bias from 
the different trigger efficiencies.
The shower trigger is sensitive not only to the neutron events but also to the photon 
events.
In order to separate the neutron events from the photon background, a cut condition
of $L_{90\%}-0.15L_{20\%} > 21$ was applied.
$L_{x\%}$ is defined by longitudinal depth
of detector where the accumulated energy deposit reaches $x\%$ of the total.
The above condition
was optimized taking into account the neutron purity and efficiency.
Events where the hadronic shower developed in the deeper GSO plates were
rejected by applying for $L_{90\%} < 37$ to improve the energy resolution of neutron.
Energy resolution of 250 GeV neutron was improved from $39\%$ to $30\%$ accordingly.

Since the RHICf detector has insufficient interaction length for the neutron
energy measurement, 
kinematic values of neutron
were unfolded by using Bayesian unfolding~\cite{bayesian}.
For prior, a Monte Carlo (MC) sample where 
neutrons from 0 to 255 GeV were uniformly generated on 
the detector was used
to avoid any bias from the particular particle productions.
We repeated the iteration until the $\chi^2$ change between two outputs of consecutive
iterations became smaller than 1.
We generated finite asymmetry by assigning up and down spin patterns to each event
and also confirmed that the unfolded distributions
reproduced the input $A_{\scriptsize{\textrm{N}}}$ well within the statistical uncertainty.

Background $A_{\scriptsize{\textrm{N}}}$s from the photon and charged hadron were
subtracted using the following equation after unfolding,
\begin{eqnarray*}
A_{\scriptsize{\textrm{N}}}^{\scriptsize{\textrm{neu}}} = 
\Big(\frac{N_{\scriptsize{\textrm{trig}}}}{N_{\scriptsize{\textrm{neu}}}}\Big)
A_{\scriptsize{\textrm{N}}}^{\scriptsize{\textrm{trig}}} - 
\Big(\frac{N_{\scriptsize{\textrm{pho}}}}{N_{\scriptsize{\textrm{neu}}}}\Big)
A_{\scriptsize{\textrm{N}}}^{\scriptsize{\textrm{pho}}} -
\Big(\frac{N_{\scriptsize{\textrm{had}}}}{N_{\scriptsize{\textrm{neu}}}}\Big)
A_{\scriptsize{\textrm{N}}}^{\scriptsize{\textrm{had}}},
\end{eqnarray*}
where the notations ``neu", ``pho", ``had" and ``trig" mean the neutron, photon,
charged hadron and triggered events, respectively, and
$N$ is the number of measured events.
Ratios between different types of events were
estimated referring to the QGSJET II-04 sample.
$A_{\scriptsize{\textrm{N}}}^{\scriptsize{\textrm{pho}}}$ was calculated using the 
photon-enhanced sample of data.
In the preliminary result, the FC and 
$A_{\scriptsize{\textrm{N}}}^{\scriptsize{\textrm{had}}}$ were not analyzed,
thereby 
$A_{\scriptsize{\textrm{N}}}$ variation between when the value of 
$A_{\scriptsize{\textrm{N}}}^{\scriptsize{\textrm{had}}}$ was $-1$ and $+1$ was assigned 
as one of the systematic uncertainties.
For additional sources of the systematic uncertainty,
uncertainties of beam center calculation, polarization and unfolding were considered.

\section{Results}
\begin{figure}[!h]
\centering
\includegraphics[width=0.6\textwidth]{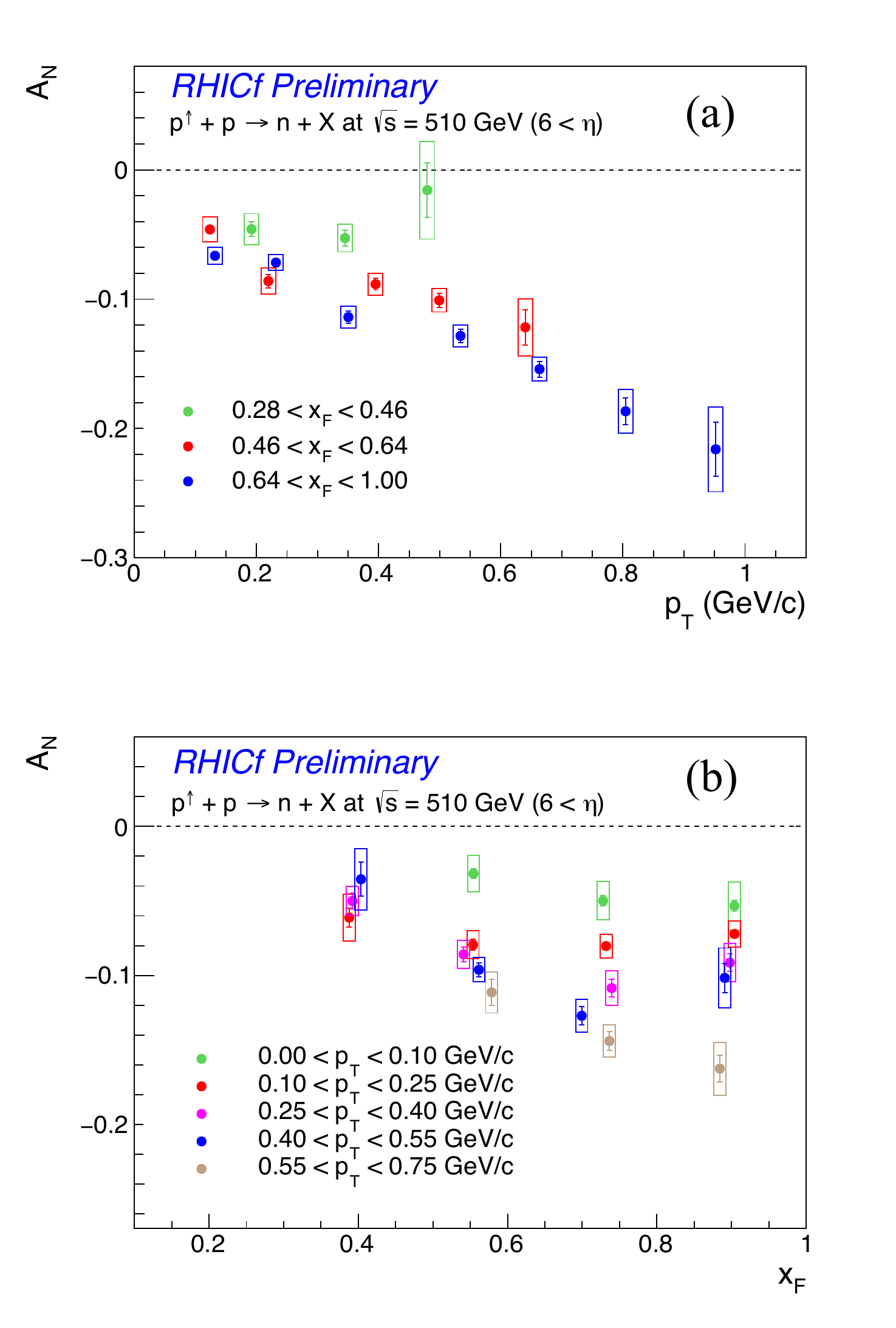}
\caption{$A_N$ for very forward neutron production (a) as a function of
$p_T$ in different $x_F$ ranges and (b) as a function of $x_F$ in
different $p_T$ ranges.}
\label{fig:preliminary}
\end{figure}

\quad \ \ 
Fig. \ref{fig:preliminary} shows the preliminary result of the $A_{\scriptsize{\textrm{N}}}$
for very forward neutron production.
Fig. \ref{fig:preliminary} (a) shows the neutron $A_{\scriptsize{\textrm{N}}}$ as a function
of $p_{\scriptsize{\textrm{T}}}$ in different $x_{\scriptsize{\textrm{F}}}$ ranges.
In the two higher $x_{\scriptsize{\textrm{F}}}$ ranges,
the $A_{\scriptsize{\textrm{N}}}$ increases in magnitude with 
$p_{\scriptsize{\textrm{T}}}$ without $x_{\scriptsize{\textrm{F}}}$ dependence
as the model predicted.
In $p_{\scriptsize{\textrm{T}}} < 0.25$ GeV/$c$,
they are consistent with those of PHENIX \cite{phenix}.
No $\sqrt{s}$ dependence was observed in the neutron $A_{\scriptsize{\textrm{N}}}$.
$A_{\scriptsize{\textrm{N}}}$s in the lowest $x_{\scriptsize{\textrm{F}}}$
range 
show different behavior compared to those in the higher $x_{\scriptsize{\textrm{F}}}$
ranges, which indicates a $x_{\scriptsize{\textrm{F}}}$ dependence that hasn't
been expected by the $\pi$ and $a_1$ exchange model.
Fig. \ref{fig:preliminary} (b) shows the neutron $A_{\scriptsize{\textrm{N}}}$ as a function
of $x_{\scriptsize{\textrm{F}}}$ in different $p_{\scriptsize{\textrm{T}}}$ ranges.
In $p_{\scriptsize{\textrm{T}}} < 0.25$ GeV/$c$, $A_{\scriptsize{\textrm{N}}}$s are
flat without $x_{\scriptsize{\textrm{F}}}$ dependence.
However, in $p_{\scriptsize{\textrm{T}}} > 0.25$ GeV/$c$, 
clear $x_{\scriptsize{\textrm{F}}}$ dependence is observed.

\begin{figure}[!h]
\centering
\includegraphics[width=0.47\textwidth]{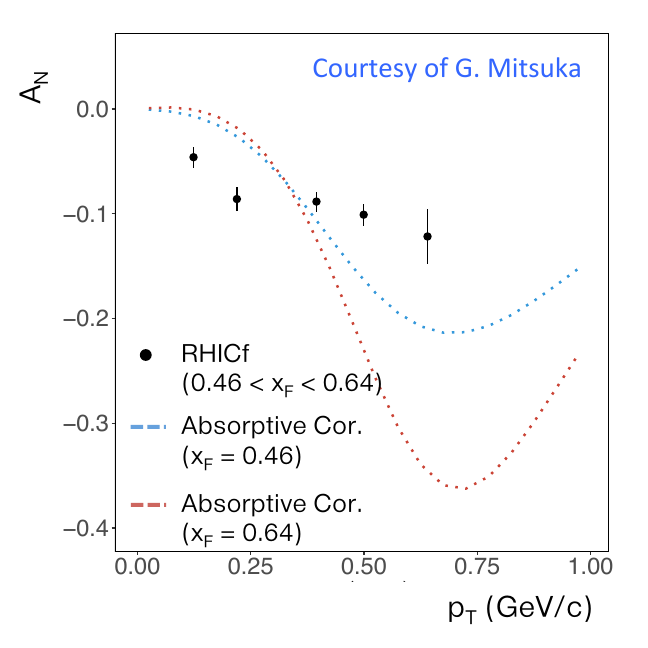}
\caption{Comparison between the RHICf data and prediction by the absorptive
correction in the single $\pi$ exchange in 
$0.46 < x_{\scriptsize{\textrm{F}}} < 0.64$.}
\label{fig:mitsuka-san}
\end{figure}

In order to understand the $x_{\scriptsize{\textrm{F}}}$ dependence of the neutron
$A_{\scriptsize{\textrm{N}}}$, spin effect by an absorptive correction in the single
$\pi$ exchange was studied.
The absorptive correction is an elastic interaction from the initial state proton or final state
multi particles.
Since phase shift by the absorptive correction starts to deviate from 
$p_{\scriptsize{\textrm{T}}} \sim 0.3$ GeV/$c$, it 
has large $A_{\scriptsize{\textrm{N}}}$
in the higher $p_{\scriptsize{\textrm{T}}}$ region.
Fig. \ref{fig:mitsuka-san} shows a comparison between the RHICf data and prediction
by the absorptive correction in $0.46 < x_{\scriptsize{\textrm{F}}} < 0.64$.
The absorptive correction doesn't reproduce the RHICf data well.
However, it shows a possible origin of the $x_{\scriptsize{\textrm{F}}}$ dependence.
Since the finite neutron $A_{\scriptsize{\textrm{N}}}$ also comes from the
$\pi$ and $a_1$ exchange, we expect that more comprehensive theoretical
calculation could reproduce the RHICf data.

To finalize the result, 
we analyzed the FC and the charged hadron background was suppressed by
by rejecting the events where ADC
of the FC has finite value.
ADC distribution of the FC was reproduced in the simulation and effect of the charged
hadron contamination after applying for the above condition was estimated.
Since the background study has been recently completed, the final result of the
neutron $A_{\scriptsize{\textrm{N}}}$ will be released soon.

\section{Summary}
In June 2017, the RHICf experiment measured the $A_{\scriptsize{\textrm{N}}}$ 
for very forward neutron production
in polarized $p + p$ collisions at $\sqrt{s} = 510$ GeV
and presented the preliminary result.
In $0.46 < x_{\scriptsize{\textrm{F}}}$, the $A_{\scriptsize{\textrm{N}}}$ increases
in magnitude with $p_{\scriptsize{\textrm{T}}}$ as the model predicted.
No $\sqrt{s}$ was observed when the RHICf data points 
were compared with those of PHENIX result.
In $p_{\scriptsize{\textrm{T}}} < 0.25$ GeV/$c$, $A_{\scriptsize{\textrm{N}}}$s are
flat without $x_{\scriptsize{\textrm{F}}}$ dependence. 
However, in $p_{\scriptsize{\textrm{T}}} > 0.25$ GeV/$c$, a clear 
$x_{\scriptsize{\textrm{F}}}$ dependence was observed.
In the preliminary result, effect of the charged hadron events was not analyzed
in detail. However, since it has been complemented, we will finalize the result soon.


\end{document}